\documentclass[prl,twocolumn,amssymb,showpacs,superscriptaddress]{revtex4}
\usepackage{graphicx,amsmath,epstopdf}

\begin{document}
\title{Emergent heavy fermion behavior at the Wigner-Mott transition}
\author{Jaime Merino}
\affiliation{Departamento de F\'isica Te\'orica de la Materia Condensada, Condensed Matter Physics Center (IFIMAC)
and Instituto Nicol\'as Cabrera,
Universidad Aut\'onoma de Madrid, Madrid 28049, Spain}
\author{Arnaud Ralko}
\affiliation{Institut N\'eel-CNRS and Universit\'e Joseph Fourier,
Bo\^ite Postale 166, F-38042 Grenoble Cedex 9, France}
\author{Simone Fratini}
\affiliation{Institut N\'eel-CNRS and Universit\'e Joseph Fourier,
Bo\^ite Postale 166, F-38042 Grenoble Cedex 9, France}
\date{\today}
\begin{abstract}
We study charge ordering driven by Coulomb interactions on triangular lattices
relevant to the Wigner-Mott transition in two dimensions.  Dynamical mean-field theory 
reveals the pinball liquid phase, a charge ordered metallic phase containing 
quasi-localized (pins) coexisting with itinerant (balls) electrons. Based on an effective periodic Anderson model
for this phase, we find an antiferromagnetic Kondo coupling between pins and balls 
and strong quasiparticle renormalization.  Non-Fermi liquid behavior 
can occur in such charge ordered systems due to spin-flip scattering of itinerant electrons off the
pins in analogy with  heavy fermion compounds.
\end{abstract}

\pacs{71.10.Hf, 73.20.Qt, 71.30.+h, 74.70.Kn}

\maketitle

The interplay between charge ordering (CO) and the Mott metal-insulator
transition (MIT) 
%is an important open problem in correlated materials, whose manifestations 
%w10
has common manifestations in narrow-band systems such as the transition metal
oxides \cite{Imada98} (cuprates, manganites, nickelates, cobaltates),
low-dimensional organic conductors \cite{SeoReview} as well as
transition-metal dichalcogenides \cite{Tosatti,Perfetti,Sipos}.  On one hand,
charge ordering phenomena are favored by the presence of electronic
correlations, as these slow down the electronic motion and reduce the tendency
to charge delocalization.
Conversely, the reconstruction of the electronic spectrum upon entering a phase
with broken symmetry can enhance the effects of correlations, extending the
realm of Mott physics
away from the usual integer fillings
\cite{Pietig,MerinoCDMFT,Pankov,Camjayi,Li,SeoReview,Amaricci10,Cano10,Cano11}.
%An illustration of this phenomenon is realized in the dichalcogenide compound
%1T-TaS$_2$ \cite{Tosatti,Perfetti,Sipos}.
%w12
% where nominally weak
%electronic correlations are so effectively enhanced upon formation of a charge
%density wave that they can  trigger a  Mott metal-insulator transition.
%w24
A fundamental relation between
charge ordering 
and the Mott MIT has also been  
%metal-insulator transition
%w2
pointed out recently in the 
%different context of the
%w4
 two-dimensional electron gas (2DEG), leading to the concept of the
Wigner-Mott transition  \cite{Pankov,Camjayi,Li,Amaricci10}.
Close to the Wigner crystal phase induced by 
the long ranged Coulomb repulsion at low density, 
the electrons form an effectively narrow-band system 
at half-filling (one electron per site of the 
Wigner crystal \cite{Lenac}), whose physical properties 
%therefore 
%w1 
resemble those of a Mott 
insulator. 
Correspondingly, it has been shown that 
the  Wigner crystal melting 
%of the
%w2
shares several common features 
with the Mott transition \cite{Pankov,Camjayi,Kotliar2012}.

In the present Letter, we explore the interplay of charge order and electronic
correlations starting from the extended Hubbard model on the  triangular
lattice, which frustrates the insulating CO found in bipartite lattices.  
We focus on the 
{\it pinball liquid} (PL) phase \cite{Kaneko06,Hotta06,Hotta-dens}, i.e.  
a charge ordered metallic phase where quasi-localized electrons (pins) coexist 
with itinerant electrons (balls)
%, and that is stabilized by the nearest-neighbor Coulomb repulsion
over a broad range of concentrations.
Applying Dynamical Mean Field Theory (DMFT)
% in order to retain both the charge and
%spin correlations, 
%w10
we find that the electron dynamics become heavily renormalized
for sufficiently large carrier concentration.
The origin of 
this renormalization is elucidated via a mapping to a 
periodic Anderson model (PAM).
%tuning the hybridization 
%between the two species reveals an antiferromagnetic coupling 
%leading to Kondo screening of the localized moments.
%Although our calculations are limited to the zero temperature case, 
%w29
The strong analogies with heavy fermions \cite{Gegenwart} indicate that
non-Fermi liquid behavior can arise in correlated CO systems on frustrated lattices
in a broad temperature range above the Kondo breakdown temperature scale.

\paragraph{Model.}

\begin{figure}
\begin{minipage}{0.2\textwidth}
\includegraphics[clip,scale=0.42]{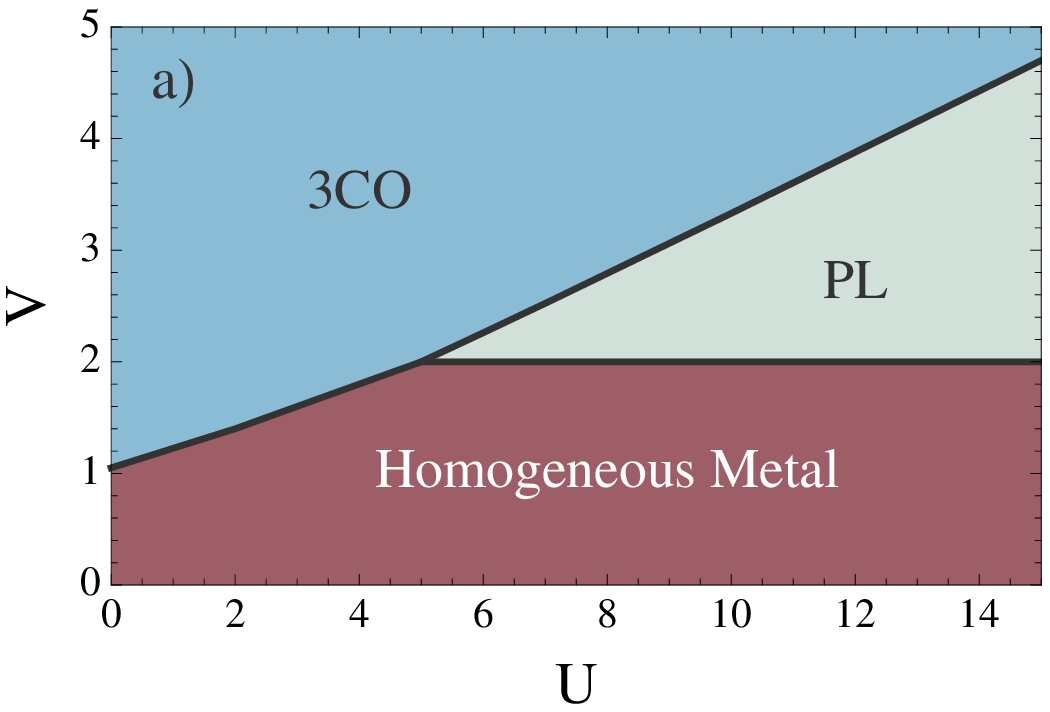}
\end{minipage}
\hspace{0.07\textwidth}
\begin{minipage}{0.2\textwidth}
\includegraphics[clip,scale=0.3]{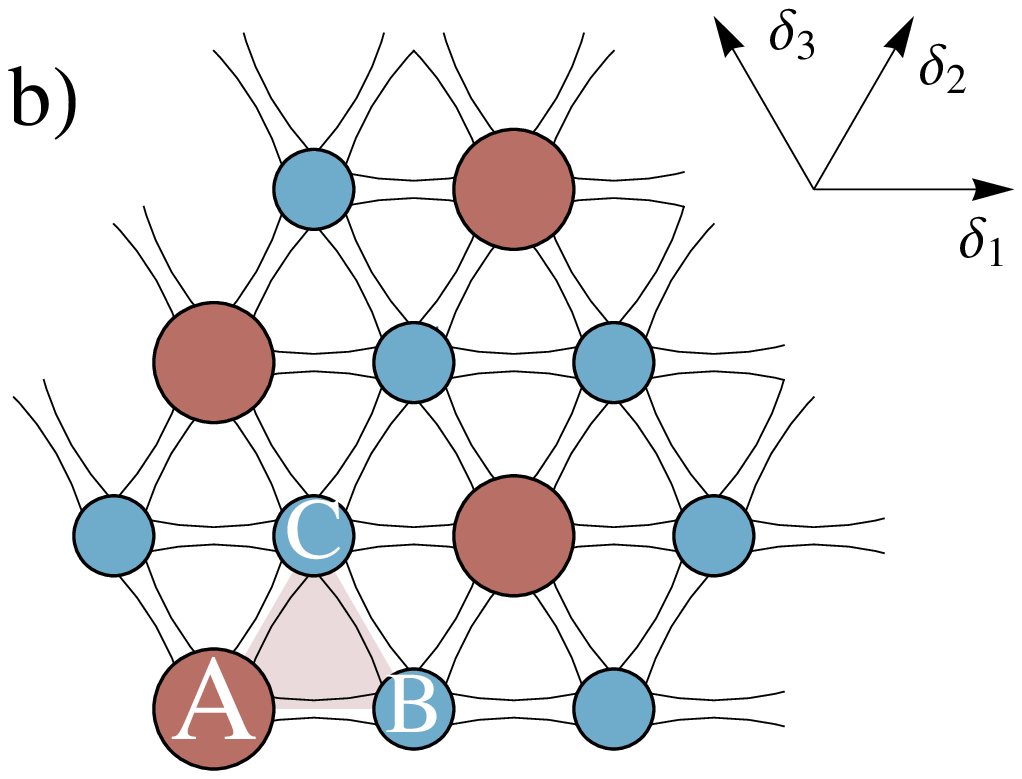}
\\
\vspace{0.06\textwidth}
\end{minipage}
\\
\includegraphics[clip,scale=0.4]{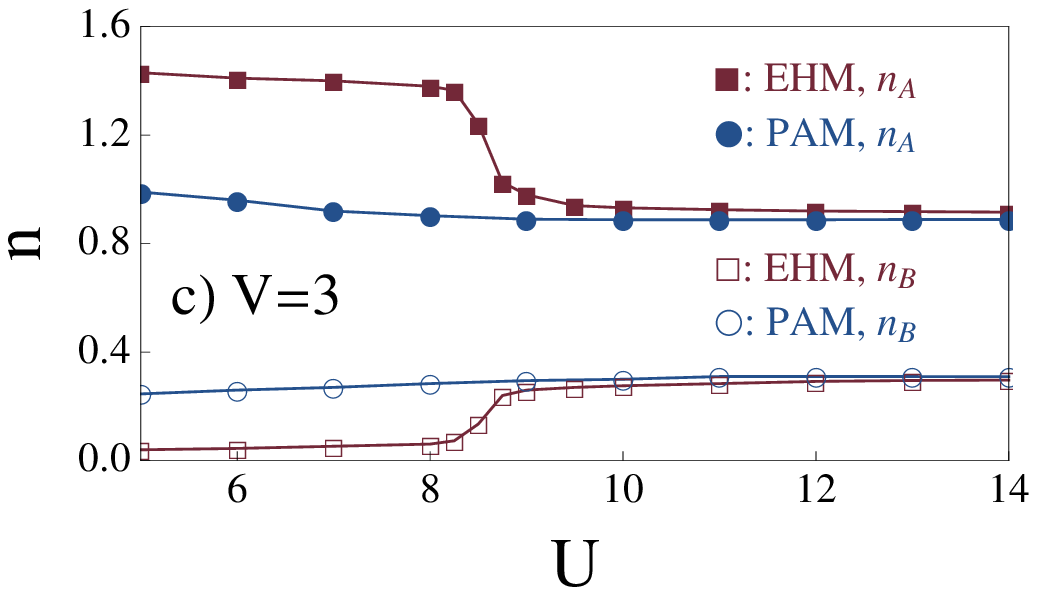}
\includegraphics[clip,scale=0.4]{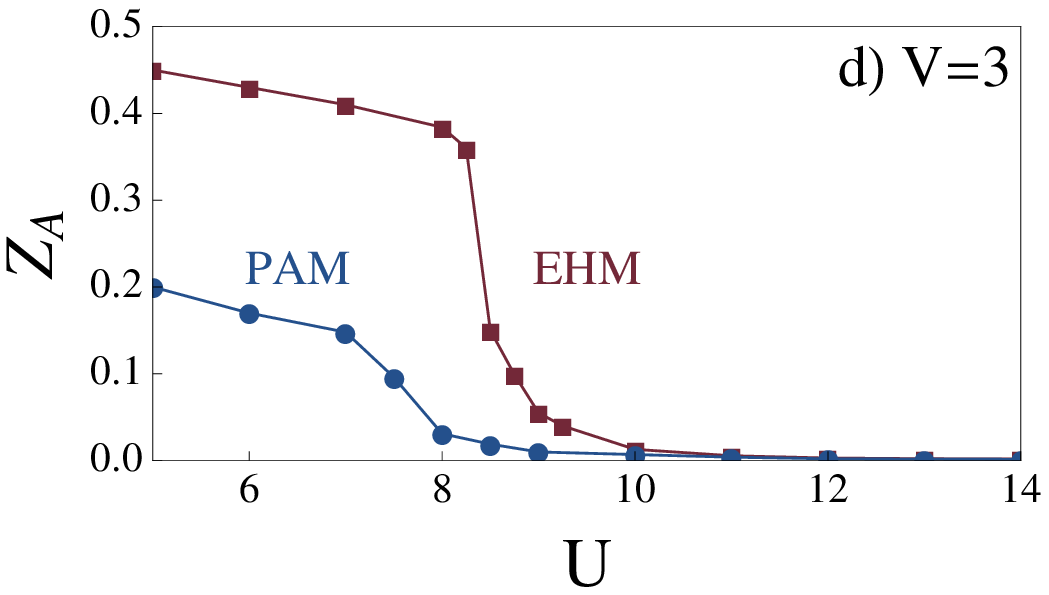}
 \caption{\label{fig:PD}(Color online)  Pinball liquid phase on the triangular lattice from DMFT.
 (a) The $U-V$ phase diagram of the EHM on the 
 triangular lattice at quarter filling. (b) Sketch of the pinball phase with pins on the $A$ sublattice
 (red, large disks) and balls (blue, small) on the $B-C$ sublattice. The shaded area is the unit cell. 
   (c) and (d)  are the charge occupations and quasiparticle weight $Z_A$ 
 as a function of $U$. Blue lines are obtained from the PAM Eq. (\ref{eq:PAM}).
}
 \end{figure}

The reference model to study 
the interplay between charge ordering and electronic correlations is the
Extended Hubbard Model (EHM): 
\begin{equation}
  \label{eq:EHM}
  H = -t\sum_{\langle ij\rangle\sigma} ( c_{i\sigma}^{\dagger} c_{j\sigma}+h.c.)
  + U \sum_i n_{i\uparrow}n_{i\downarrow}  +
   V \sum_{\langle ij\rangle}  n_i n_j.
\end{equation}
which describes electrons moving on a lattice with transfer integral $t$ and
interacting via both on-site ($U$) and nearest-neighbor  Coulomb interactions
($V$). We set the energy scale $t =1$.  Studies of the EHM have mostly focused
on bipartite lattices \cite{SeoReview,Camjayi,Pietig,Amaricci10,Li,MerinoCDMFT}
(the Bethe lattice, or the square lattice in two-dimensions), as these can
naturally accommodate for charge ordering at a commensurate occupation of one
charge per two lattice sites, corresponding to one quarter band filling.  There
are important cases however where the relevant lattice geometry is
non-bipartite, adding frustration to the already rich phenomenology of
correlated CO. A notable example is the two-dimensional Wigner crystal itself,
where charges arrange on a triangular pattern in order to minimize their
electrostatic energy. 
How the analogy between the Wigner crystal
transition and the Mott MIT is modified when one 
considers the appropriate triangular geometry   \cite{Lenac}
remains an open question, that was not addressed in Refs. \cite{Pankov,Camjayi}.
A second example is the family of  two-dimensional 
organic conductors  $\theta$-(BEDT-TTF)$_2$X \cite{Kino96,Mori98}, 
where the molecular lattice is triangular
and the band filling is fixed by charge transfer 
to one hole per two molecular sites. 
In this framework, 
the EHM has been studied intensively to address the phenomenon of charge ordering, and fundamental
differences between the square and triangular lattice
have emerged. Most prominently,
for isotropic $V$, on the triangular lattice, 
geometrical frustration prevents the two-fold CO insulator characteristic
of bipartite lattices, and a three-fold CO metal is realized instead (cf. Fig. \ref{fig:PD}b)
\cite{Kaneko06,Hotta06,Cano10,Cano11}.

\paragraph{DMFT solution.}

%We solve the model Eq. (\ref{eq:EHM}) using DMFT.
%w12
In its simplest form, DMFT maps a single-band Hubbard model onto an
Anderson quantum impurity model in an effective metallic bath that is
determined self-consistently.
Allowing for charge ordering \cite{Pietig} with $N$-fold symmetry breaking
implies solving $N$ interconnected (through hopping amplitudes) impurity
models, one for each sublattice.
Calling 
$G^0_{\alpha,\sigma}(i\omega_n)$ the Green's function of the $\alpha$-th electronic bath and 
$G^{And}_{\alpha,\sigma}(i\omega_n)$ that of the corresponding impurity problem, 
a local self-energy is extracted from   
the solution of the Anderson model for each sublattice $\alpha$ via $
\Sigma_{\alpha,\sigma}(i\omega_n)=G^0_{\alpha,\sigma}(i\omega_n)^{-1}-
G^{And}_{\alpha,\sigma}(i\omega_n)^{-1}.
$
The lattice Green's function  within the dynamical 
mean-field approximation then reads
$
{\bf G}_{\sigma}({\bf k}, i \omega_n)
=[(i \omega_n + \mu) {\bf I} - {\bf E}_H  - {\bf t}({\bf k})- 
{\bf \Sigma}_\sigma(i\omega_n) ]^{-1}, 
\label{eq:Gk}
$
where bold letters indicate $N \times N$ matrices in sublattice space,  ${\bf
I}$ is the identity matrix and ${\bf t}(k)$ is the Bloch matrix connecting the
different sublattices and $\mu$ 
is the chemical potential.  Matsubara frequencies, $\omega_n=(2n+1)\pi/\beta$, with
$\beta=200$ are used and $n_{max}=2048$. 
Specializing to the three-fold order realized on the triangular lattice, the
matrix elements between sublattices $\alpha=A,B,C$ are
$t_{AB}(k)=t_{BC}(k)=\phi_k$, $t_{AC}(k)=\phi_k^*$, having defined
$\phi_k=\sum_i e^{i\delta_i\cdot k}$ and $\delta_i$ ($i=1,3$) the vectors
connecting an $A$ site to its three $B$ neighbors (see Fig.~\ref{fig:PD}(b)).
${\bf E}_H$ is a diagonal matrix whose elements $E_{H,\alpha}=3\sum_{\gamma \ne
\alpha} V \langle n_\gamma \rangle $ are, for each $\alpha$, the mean-field
electrostatic potentials of charges on the nearest-neighboring sites.
The problem is solved self-consistently by imposing the condition $(1/L)
\sum_{\bf k} G_{\alpha\alpha,\sigma}({\bf k}, i \omega_n) =
G_{\alpha,\sigma}^{And}(i\omega_n)$, with $L$ the total number of lattice
sites. 
We use Lanczos diagonalization as a problem solver, taking $N_s=10$ sites
for each quantum impurity problem. 
\cite{note1}  
The spectral density 
of electrons on the $\alpha$-sublattice is obtained from: 
$A_{\alpha,\sigma}(\omega)$=-Im$G_{\alpha\alpha,\sigma}(\omega+i\eta)/\pi$, 
with $\eta=0.15$.
We restrict our analysis to solutions where the $B$ and $C$ sublattices 
are equivalent.
Practical calculations for a density $n$ of holes are performed by taking  $n$
electrons per site and changing the sign of $t$ in Eq.~(1).

\paragraph{The pinball liquid.}

\begin{figure}
\includegraphics[clip,scale=0.65]{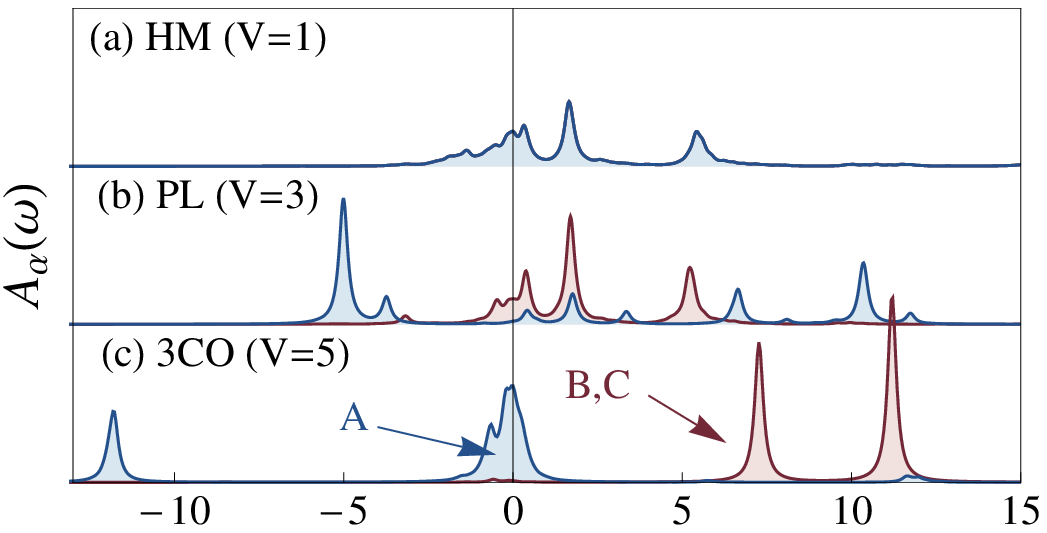}
\\
\vspace{-0.02\textwidth}
\includegraphics[clip,scale=0.65]{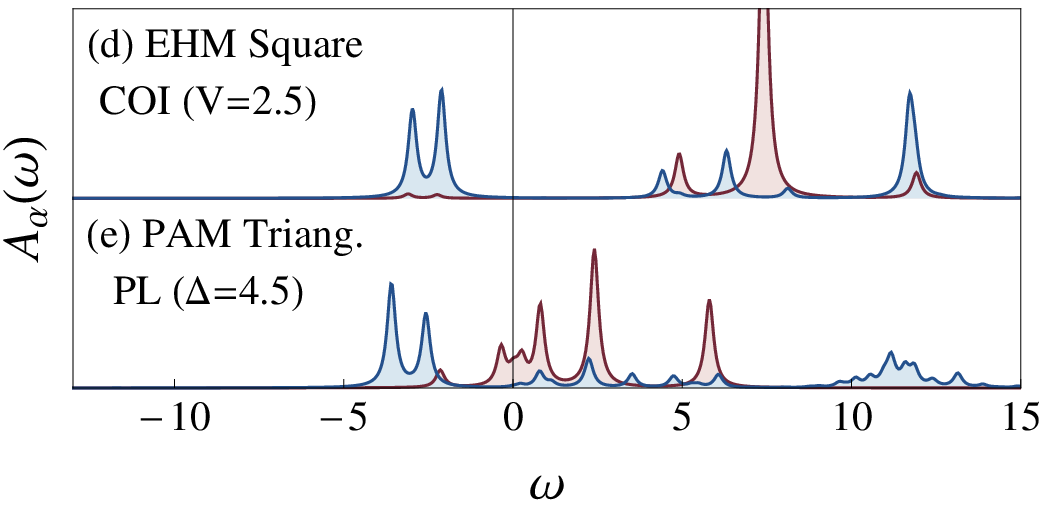}
 \caption{\label{fig:spec}(Color online) 
Spectral densities
$A_A(\omega)$ and $A_B(\omega)$ projected on the charge rich and charge poor
sublattices, for $U=12$ in the EHM on the triangular lattice.  
(a) HM, (b) PL and (c) 3CO phase. (d) shows the same quantity in the
CO insulating phase on the square lattice. (e) 
is the spectral density calculated for the PAM at $\Delta=4.5$. 
%A small imaginary part $\delta=0.15t$ has been used.
%w10  but they ask for it!
The vertical line denotes the
Fermi energy. 
}
\end{figure}

Fig.~\ref{fig:PD}(a) shows the phase diagram of the EHM
obtained from DMFT on the triangular lattice at $n=1/2$ (quarter filling), 
which basically agrees with previous results obtained from exact
diagonalization (ED) on small clusters \cite{Cano10,Cano11}.
Figs. \ref{fig:spec} (a), (b) and (c) 
show representative spectral densities obtained in 
the three different phases.
A transition from a homogeneous metal (HM)
to a metallic CO state (3CO)
is realized in the weakly correlated limit, $U\to 0$, upon increasing the intersite repulsion
$V$, similar to what is found on bipartite lattices
\cite{Camjayi,Amaricci10,Li,MerinoCDMFT}. A large gap opens up in the spectrum due to  
the CO phenomenon (Fig. \ref{fig:spec}(c)), but the Fermi energy does not fall in the gap 
and the system remains metallic.
At large $U$, instead, a different ordered metallic phase emerges, 
termed the {\it pinball liquid} \cite{Kaneko06,Hotta06}, extending
up to increasingly large values of  $V$.
In such  phase, quasi-localized charges  ({\it pins}) coexist with  itinerant
carriers ({\it balls}) flowing in the honeycomb lattice formed by interstitial
sites (see Fig.~\ref{fig:PD}(b)). Unlike the 3CO phase, where conduction arises 
from the charge rich sites, in the PL it is  the minority  
carriers (balls) who are responsible for the metallic behavior 
(see Fig.~\ref{fig:spec}(b)). The corresponding conduction band has a large width 
(comparable  with that of the homogeneous metal itself) at all values of $V$, and does not
scale with $\propto t^2/V$ as in the 3CO phase. Note that the pinball phase is absent on 
the square lattice, where the system is CO and {\it insulating} (COI) at 
sufficiently large $U$ \cite{Camjayi,Amaricci10,Li,MerinoCDMFT},  cf.  Fig.~\ref{fig:spec}(d).

The origin of the pinball phase can be best understood as follows.
At large $V$ and small $U$, the 3CO phase has
a density $n_A\to 3/2$ on the charge-rich sites and $n_B\to 0$ on the
charge-poor sites. 
Increasing $U$ tends to suppress the double occupancy and
eventually when $U\gtrsim 3V$, the local Coulomb energy cost is avoided  by
leaving only one charge per $A$ site (a pin, $n_A\simeq 1$).  The excess charge
density is then transferred to the charge-poor sites (balls, $n_B\simeq 1/4$).
This transition, shown in Fig.~\ref{fig:PD}(c), occurs between two {\it metallic} phases:
which only differ on the amount of charge transferred. This transition is of a different type to the
metal-to-COI transition \cite{Amaricci10} on the square lattice, which displays phase coexistence 
as in the pure Mott metal insulator transition.
We  find that the two charge species behave very differently: pins form an almost
half-filled system with a marked Mott behavior due to strong onsite Coulomb
interactions, in opposition to the balls which, owing to their low concentration,
are in principle protected from them. The spectral function for the $B,C$ sublattices
indeed shows a wide band crossing the Fermi energy, while
the  $A$-sublattice shows typical features of strongly correlated systems: the
spectral weight is mostly located in two separate regions away from the Fermi level -- the 
lower and upper Hubbard bands separated by an energy $U$ --
while the weight at the Fermi energy is strongly suppressed.
This can be quantified via the quasiparticle weight of the correlated pins,
 $Z_A=\left[ 1-\partial Im \Sigma_A(\omega)/\partial \omega \right]^{-1}_{\omega=0}$, 
which is depicted in Fig.~\ref{fig:PD}(d). It
is large in the conventional 3CO phase at low $U$, and drops drastically upon
entering the pinball liquid (here at $U=8.5$).
This drop occurs when the $A$-sublattice concentration spontaneously adjusts to $n_A
\simeq 1$ (Fig.~\ref{fig:PD}(c)). Within the accuracy of our method, a  small non-vanishing 
quasiparticle weight persists at all $U$ 
\cite{note2}
Since the filling of the correlated hybrid band is never
exactly an integer, the results indicate that 
the pins do not undergo a full Mott transition, but remain in a strongly correlated Fermi liquid state.
This contrasts with the situation encountered on bipartite lattices
\cite{Amaricci10}, where the electron concentration in the correlated band is
automatically fixed to half-filling once that the Fermi energy falls inside the
gap, resulting in a true Mott transition for $U\gtrsim W$ (the bandwidth) and $V \gtrsim 2 $.

\paragraph{Periodic Anderson model and Emergent heavy fermions.}
To get further insight into the effects of electronic correlations we now
map the problem into a  periodic Anderson model (PAM). 
This is achieved by introducing creation operators $a^+_i$, $b^+_i$ and $c^+_i$
respectively on the $A$, $B$ and $C$ sublattices and diagonalizing the bands of
the honeycomb lattice formed by $B$ and $C$ sites (tight-binding on graphene).
Rewriting $\phi_k=\sum_i e^{i\delta_i\cdot k}$ as $E_k e^{i\theta_k}$, we can
define $B$-$C$ subspace spinor operators
$\gamma^+_{k,\pm}=(e^{-i\theta_k/2},\pm e^{i\theta_k/2})/\sqrt{2}$, whose
energies are $\epsilon_{k,\pm}=\pm E_k$ (the origin is fixed at the Dirac
point). Dropping the Coulomb interaction on the conduction bands leads
to
\begin{eqnarray}
H_{\textrm{PAM}} &=& \sum_{i\sigma} \epsilon_A a^\dagger_{i\sigma} a_{i\sigma} 
+  \sum_{k\sigma\alpha} \epsilon_{k,\alpha} \gamma^+_{k\alpha\sigma}\gamma_{k\alpha\sigma}
\\ &+& \sum_{ik,\alpha} (V_{k,\alpha} e^{ikR_i} a^\dagger_{i\sigma}\gamma_{k\alpha\sigma} + h.c.)
+ U \sum_i a^\dagger_{i \uparrow} a_{i\uparrow} a^\dagger_{i \downarrow} a_{i\downarrow}  
\nonumber \label{eq:PAM}
\end{eqnarray}
where the hybridization between the conduction bands and the localized level 
is determined by 
$V_{k,-}=i\sqrt{2} E_k \sin(3 \theta_k/2)$ and $V_{k,+}=\sqrt{2} E_k \cos(3 \theta_k/2)$.
The energy of the localized level is set by the charge transfer gap,
$\epsilon_A=-\Delta$, which originates from the electrostatic interaction
between charges on different sublattices. 
The pinball phase 
%at $n=1/2$ 
studied in the preceding paragraphs corresponds to $\Delta=3V/2$ and is therefore 
in the charge transfer regime $U\gg \Delta$ (because $U>3V\gg 3V/2$, cf. 
Fig.~\ref{fig:PD}(a)).
The quantities $n_\alpha$, $Z_A$ and $A_{\alpha}(\omega)$ calculated in the
PAM are compared with the EHM results in Figs.~\ref{fig:PD}-(c,d) and \ref{fig:spec}-(a,b)
showing a consistent agreement in the pinball liquid phase. 

The nature of the coupling between pins and balls is now conveniently analyzed 
based on the low energy effective PAM, varying the number of electrons per unit cell $n_{\textrm{cell}}
\in [0,3]$ ($n_{\textrm{cell}}=3n$) for a fixed $\Delta=4.5$.
This interval includes the region of stability of the pinball in the original
EHM \cite{Hotta-dens}, i.e. $1 < n_{\textrm{cell}} < 2$.
\begin{figure}
\includegraphics[clip,scale=0.65]{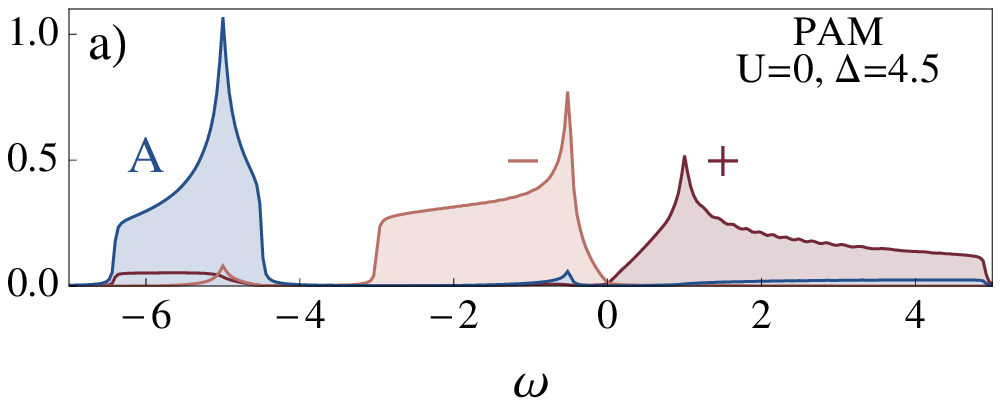}
\\
\includegraphics[clip,scale=0.34]{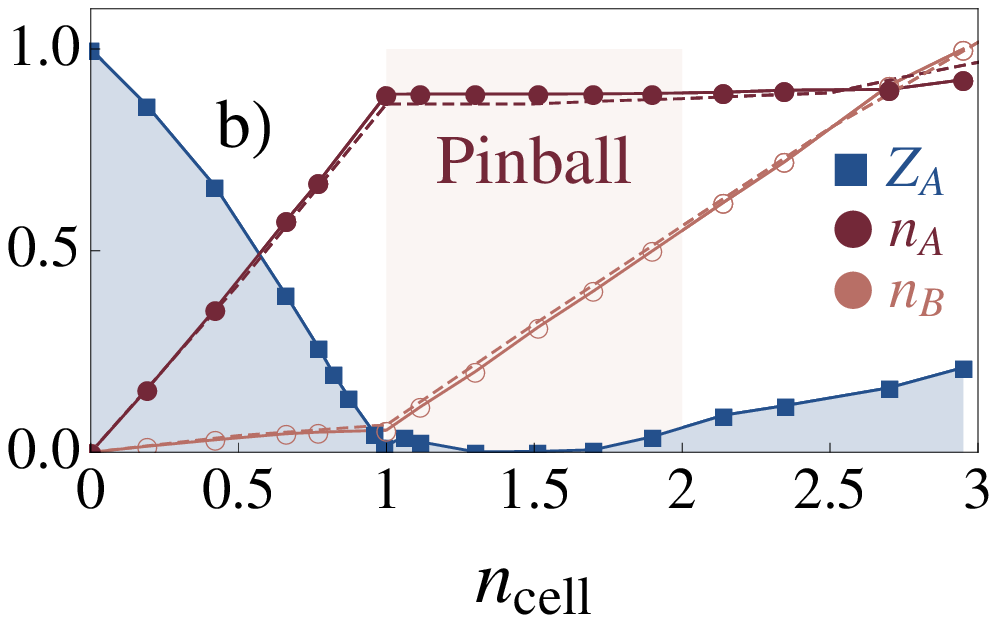}
\includegraphics[clip,scale=0.34]{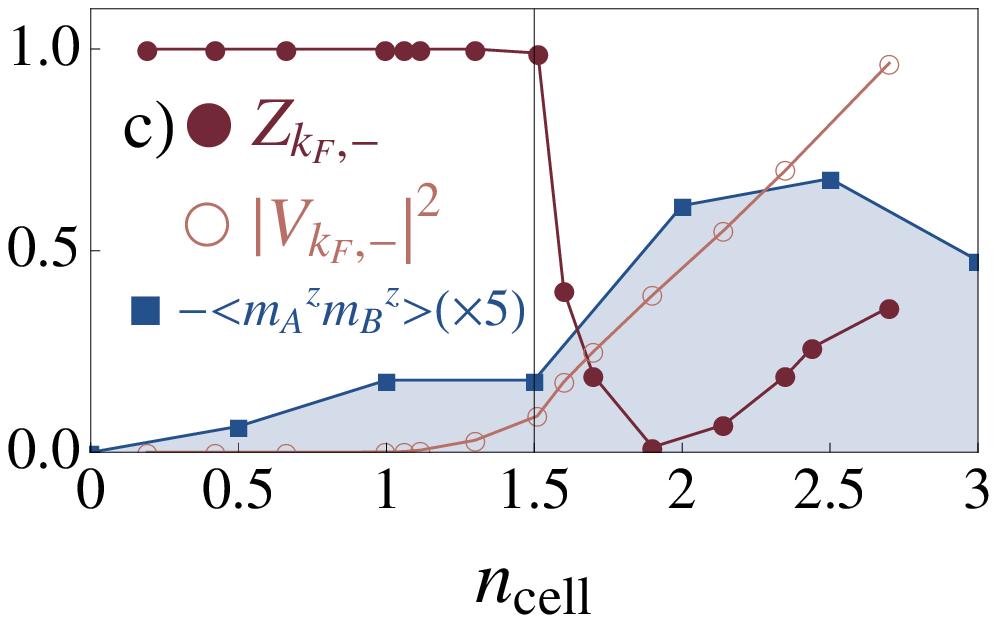}
\caption{\label{fig:PAM}(Color online)  Strong quasiparticle renormalization 
in pinball liquid phase and antiferromagnetic Kondo correlations.  
(a) DOS of the noninteracting PAM ($U=0$)  for 
$\Delta=4.5$. Different colors correspond to projection on the non-hybridized bands: pins (blue),
$-$ (orange) and $+$ (red). (b) A/B sublattice densities  (full lines=DMFT, dashed=ED) 
and pin quasiparticle weight (DMFT) and 
(c) hybridization, quasiparticle renormalization (DMFT) and magnetic correlations (ED) as a function of the filling  
$n_{cell}$ at $U=12$.
}
\end{figure}
The density of states (DOS) of $H_{\textrm{PAM}}$ in the non-interacting limit
$U=0$ is shown in Fig.~\ref{fig:PAM}(a).  Hybridization
broadens the localized pin level originally at $\epsilon_A=-\Delta$ 
 into a dispersive band, and also affects the dispersion
of the conduction bands, which is no longer symmetric around the Dirac point $\omega=0$.
Because $\epsilon_A < 0$, pins are populated first and $n_A\simeq
n_{\textrm{cell}}$ until $n_{\textrm{cell}}=1$ as  depicted
in Fig.~\ref{fig:PAM}(b). 
For $n_{\textrm{cell}}> 1$, the $A$ sublattice is filled with $\simeq 1$
electron per site (the maximum allowed by the large $U$) and the remaining
electrons move to the conduction band. Correspondingly, the Fermi energy moves
from the bottom of the $(-)$ band to the Dirac point, reached at $n_{\textrm{cell}}=3$. 
Interestingly,  $Z_A$ (blue squares in Fig.~\ref{fig:PAM}(b)) is strongly suppressed in the whole 
pinball phase  (hatched region in Fig.~\ref{fig:PAM}(b)).
Following Refs. \cite{Hewson,Amaricci12}, we define the effective self-energy
for conduction electrons as
\begin{equation}
\Sigma_{k,\pm}(i \omega_n)=\frac{|V_{k,\pm}|^2}{i\omega_n+\mu-\epsilon_A-\Sigma_A(i\omega_n)-
\frac{|V_{k,\mp}|^2}{i\omega_n+\mu-E_k}}
\label{eq:sigma-cond}
\end{equation}
which accounts for hybridization, and includes crucial interaction effects
arising indirectly via the Coulomb repulsion of localized charges. 
Based on Eq. (\ref{eq:sigma-cond})  we evaluate the quasiparticle
renormalization $Z_{k,\pm}=[1-\partial Im (\Sigma_{k,\pm})/\partial
\omega]^{-1}_{\omega=0}$ of the balls.  Note that $V_{k,-}$ vanishes for wavevectors $k$
forming an angle $\pi/6 \ [\pi/3]$ with the x-axis, leading to {\it cold} spots where conduction
and localized electrons are perfectly decoupled and $Z_{k,-}=1$ identically.
In Fig.~\ref{fig:PAM}(c) we report instead $Z_{k,-}$ in the directions of
maximum hybridization on the Fermi surface ({\it hot} spots), for which a sharp
drop is found at $n_{\textrm{cell}}=3/2$. This is  attributed to an increase in
the hybridization with the correlated pins occurring around this filling (see
also $|V_{k,-}|^2$).
On the same figure is also reported the inter-species magnetic correlation
$\langle m_z^A m_z^B\rangle=\langle (n_{A\uparrow} - n_{A\downarrow})
(n_{B\uparrow} - n_{B\downarrow})\rangle$ calculated by ED on a 12-site cluster
(we use ED because inter-site correlations cannot be addressed in the single site DMFT scheme).
As shown in Fig.~\ref{fig:PAM}(c), the strong renormalization of the metal is accompanied by a
buildup of anti-ferromagnetic correlations, corresponding to a screening of the
localized moments by the conduction electrons.
From our data we conclude that pins and balls are nearly decoupled for
$n_{\textrm{cell}}\le 3/2$,  while 
a strong Kondo coupling between the two fluids, accompanied by
a strong renormalization of the conduction electrons, arises at larger concentrations.

\paragraph{Wigner crystal in the 2DEG.}
According to the lattice analogy  proposed in Refs.
\cite{Lenac,Pankov,Camjayi}, the existence of pronounced short range order
close to the MIT enables a mapping of the continuous 2DEG  to our model
$H_{\textrm{PAM}}$ at $n_{\textrm{cell}}=1$. We can then associate $\Delta \propto r_s$ \cite{Lenac}, 
where $r_s$ is 
the ratio of Coulomb to kinetic energy in the electron gas.
The case $\Delta=4.5$ analyzed in Fig.~\ref{fig:PAM}(b) then corresponds,
for $n_{\textrm{cell}}=1$,
to a 2DEG well into the Wigner crystal phase: 
the system is insulating, as both $Z_A$ and the DOS at the Fermi 
energy (not shown) vanish at this point.  
Following Ref. \cite{Lenac} the  quantum melting of the Wigner crystal upon reducing $r_s$ 
can be identified with  a closing of the
gap between the pin and $(-)$ bands (Fig.~\ref{fig:PAM}(a)) occurring below a critical $\Delta_c$. 
From our DMFT results, 
this happens at $\Delta_c\simeq 3.7$ for $U=12$. 
Although a detailed study of the MIT of the 2DEG is beyond the scope of this work, 
we note that a metallic state can also be achieved at $\Delta>\Delta_c$, by doping away from 
$n_{\textrm{cell}} = 1$.
The resulting metal is very
different when doping  holes into the pin band ($n_{\textrm{cell}} = 1+\delta$ with $\delta<0$, 
leading to $Z_A\propto
|\delta| $, cf. Fig.~\ref{fig:PAM}(b)) or electrons into the conduction band ($\delta>0$, with 
$Z_A$ jumping sharply to a small but finite value).
The latter situation actually corresponds to the self-doping instability proposed in
Ref.  \cite{Pankov}. If this scenario were realized,
the metal on the verge of Wigner crystallization would actually 
be a pinball liquid itself, 
indicating an appealing connection between
the physics of the 2DEG and the emergence of heavy fermion behavior.

\paragraph{Concluding remarks.}

The combination of electronic correlations and 
geometrical frustration on charge ordered triangular lattices leads to an 
exotic pinball liquid state in which localized and itinerant electrons coexist.
Based on the close analogies with heavy fermion systems \cite{Gegenwart}, pointed out via an
explicit mapping of the electronic problem onto the periodic Anderson model, 
non-Fermi liquid behavior should occur in such phase as a consequence of 
the scattering of the itinerant electrons off the localized moments of the pins. 
%On the experimental side, this behavior would be consistent with the anomalous 
%metallic phases of the family of organic 
%conductors $\theta$-(BEDT-TTF)$_2$X which display violation of the Mott-Ioffe-Regel limit  in the
%resistivity,  absence of Drude peak in the optical conductivity \cite{takenaka2005,Tajima2000} and 
%non-monotonic 1/T$_1$T-NMR relaxation rates \cite{Kanoda2012} close to charge ordering. T
The scenario proposed here could be of relevance to several  
 correlated materials exhibiting charge ordering on triangular lattices
 such as the organic conductors  $\theta$-(BEDT-TTF)$_2$X \cite{Kanoda2012},  transition-metal dichalcogenides 
 \cite{Tosatti,Perfetti,Sipos}, AgNiO$_2$ \cite{Coldea2007} and
 $^3$He bilayers \cite{Neumann}, as well as to the interaction-driven MIT in the 2DEG 
 \cite{Pankov,Camjayi,Amaricci10}. 
%w52 117 instead of 169

\section*{Acknowledgments.} 
J.M. acknowledges financial support from MINECO (MAT2012-37263-C02-01).  This work is supported by the French National Research
Agency through Grant No. ANR-12-JS04-0003-01 SUBRISSYME.


\begin{thebibliography}{30}
\bibitem{note1}This is sufficient to find a reliable projection to the Anderson model 
through conjugate gradient minimization of: 
$\sum_n |G^0_{\alpha\sigma}(i\omega_n)-G^{0And}_{\alpha\sigma}(i\omega_n)|$, 
for each $\alpha$-sublattice.
\bibitem{note2}
Due to finite discretization, a small $Z_\alpha$ obtained from the smallest Matsubara 
frequency could be consistent with a Mott insulator\cite{Bluemer}.
\end{thebibliography}
\end{document}